\documentclass[prb,twocolumn,showpacs,superscriptaddress]{revtex4}
\usepackage[dvips]{graphicx}
\usepackage{color}
\usepackage{latexsym}
\usepackage{amsmath}
\usepackage{amssymb}

\newcommand{\Ad}[1]{{\mathbf a}^\dag_{#1} }
\newcommand{\A}[1]{{\mathbf a}_{#1} }

\newcommand{\expect}[1]{\langle \Phi | #1 | \Phi \rangle}

\begin{document}
%
%
%
%
\preprint{LA-UR-04-6662}
\title
   {BCS-BEC crossover with a finite-range interaction}

\author{Meera M. Parish}
\email{mmp24@cam.ac.uk} \affiliation{Cavendish Laboratory,
Madingley Road, Cambridge CB3 0HE, United Kingdom}

\author{Bogdan Mihaila}
\email{bmihaila@lanl.gov} \affiliation{Theoretical Division, Los
Alamos National Laboratory, Los Alamos, NM 87545}

\author{Eddy M. Timmermans}
\email{eddy@lanl.gov} \affiliation{Theoretical Division, Los
Alamos National Laboratory, Los Alamos, NM 87545}

\author{Krastan B. Blagoev}
\email{krastan@lanl.gov} \affiliation{Theoretical Division, Los
Alamos National Laboratory, Los Alamos, NM 87545}

\author{Peter B. Littlewood}
\email{pbl@cam.ac.uk} \affiliation{Cavendish Laboratory, Madingley
Road, Cambridge CB3 0HE, United Kingdom} \affiliation{National
High Magnetic Field Laboratory, Los Alamos National Laboratory,
Los Alamos, NM 87545}

%
\begin{abstract}
We study the crossover from BEC to BCS pairing for dilute systems
but with a realistic finite-range interaction. We exhibit the
changes in the excitation spectrum that provide a clean
qualitative distinction between the two limits. We also study how
the dilute system converges to the results from a zero-range
pseudo-potential derived by Leggett.
\end{abstract}


%
\maketitle
%
%
\section{Introduction}


Ultracold atomic gases provide an experimental playground for
testing pairing phenomena, due to the ability to control the
inter-atomic interactions via a magnetically-tuned Feshbach
resonance \cite{ref:TVS93}. Over the past year, tremendous
progress has been made in realizing the crossover from the
Bose-Einstein condensation (BEC) of diatomic molecules to the
Bardeen-Cooper-Schrieffer (BCS) limit of weakly-bound Cooper pairs
in ultracold gases of fermionic atoms~\cite{fermi_exp,exp_04}.
But there is much debate over where the crossover between the BCS
and BEC regimes occurs, following recent experimental studies on
fermionic condensates \cite{exp_04}. Therefore, it is timely to
perform a thorough theoretical investigation of what the signature
of the crossover is and how it might be observed experimentally,
using a realistic potential with a finite range.


Much of the theoretical work on systems of two types of fermions
interacting via an adjustable, attractive potential have focussed
on interactions that are governed by a single parameter, namely
the s-wave scattering length~$a_0$ \cite{ref:leg80,ref:ran}.
Such a description is valid provided we have $|a_0|\gg \langle r
\rangle$ and $k_F \langle r \rangle \ll 1$, where $\langle r
\rangle$ is the width of the potential and $k_F$ is the Fermi
momentum of the non-interacting system, so that the only
independent dimensionless variable in the problem is $a_0 k_F$.
Thus, it is only applicable to dilute systems like the ultracold
atomic gases, not the high density situation found in conventional
superconductors \cite{ref:schr}. The potential supports a 2-body
bound state for $(a_0 k_F)^{-1}~>~0$, but this molecular state
passes through zero energy and vanishes into the continuum at
$(a_0 k_F)^{-1} = 0$, the position of the Feshbach resonance. As
such, the BCS and BEC limits correspond to $(a_0
k_F)^{-1}~\rightarrow~- \infty$ and $(a_0 k_F)^{-1}\rightarrow +
\infty$, respectively.

The distinction between the BEC and BCS regimes can be made
sharper~\cite{ref:leg80}:
the qualitative boundary between the BCS and BEC ground states
occurs when the chemical potential $\mu$ reaches $0$,
marking the disappearance of a Fermi surface, and this coincides
with the appearance of a bound state at $(a_0 k_F)^{-1} = 0$ only
when the density is zero. At the point where there is no longer a
defined Fermi surface, there are qualitative changes in the
quasi-particle excitation spectrum: the momentum corresponding to
the minimum in the excitation spectrum shifts from finite momentum
in the BCS limit to zero momentum in the BEC limit. At the same
point, the value of the excitation gap goes from $\Delta$ in the
BCS limit to $\sqrt{\mu^2 + \Delta^2}$ in the BEC limit, where
$\Delta$ is the gap parameter. We shall use these changes in the
excitation spectrum as the ``definition" of the crossover point.

There have also been studies of the BCS-BEC crossover using
Coulomb potentials in the context of excitons \cite{ref:CN82}.
Here, the interaction is kept fixed and the density varied to
achieve a smooth transition from the high density BCS limit to the
dilute BEC limit. Note that it is not possible to obtain a
density-driven crossover using an interaction that is only
characterized by the scattering length, since this picture breaks
down at high densities.

In this paper, we shall study the BCS-BEC crossover at zero
temperature using a realistic Gaussian potential and compare with
results from single-parameter potentials. We use a mean-field
variational wave function to describe the crossover, which is
expected to give an accurate description of the low temperature
behavior, though not, of course, the finite temperature
transition.

The paper is organized as follows: We describe the model and
variational scheme in Sec.~\ref{formalism}. We discuss the
characteristics of the wave functions and excitation spectra in
Sec.~\ref{spectra}. In Sec.~\ref{universal} we compare our
results with Leggett's zero-range pseudo-potential result which is
usually used to describe the mean-field limit for dilute Fermi
systems. We conclude in Sec.~\ref{conclusions}.

%
%

\section{Formalism}
\label{formalism}

We determine the ground-state wave function $| \Phi \rangle$ using
conventional mean-field methods, briefly reviewed here. We
consider the Hamiltonian
\begin{align}
   \hat{H} =
   \sum_{\mathbf{k},i} \epsilon_{\mathbf{k}} \
   a^{\dag}_{\mathbf{k} i} a_{\mathbf{k} i}
   +
    \
   \sum_{\mathbf{k,p,q}} V_{\mathbf{q}} \
                 a^{\dag}_{\mathbf{k} \uparrow}
                 a^{\dag}_{\mathbf{p} \downarrow}
                 a_{\mathbf{p} - \mathbf{q} \downarrow}
                 a_{\mathbf{k} + \mathbf{q} \uparrow}
   \>,
\label{eq:ham_0}
\end{align}
where $\mathbf{k}$ and $i$ denote momentum variables and spin
states $\{\uparrow, \downarrow \}$, respectively, and
$\epsilon_{\mathbf{k}} = \hbar^2 \mathbf{k}^2/2m$.
Since experiments on atomic gases are carried out at low energies,
we will restrict ourselves to the simple case of an \emph{s}-wave
pairing interaction.

We calculate the ground state by minimizing the free energy
\begin{align}
   \mathbf{F}
   \ = \
   \langle \Phi | \hat{H} - \mu \hat{N} | \Phi \rangle
   \>,
\end{align}
where $\hat{N} = \sum_{\mathbf{k}, i} a^{\dag}_{\mathbf{k} i}
a_{\mathbf{k} i}$ is the total number operator and $\mu$ is the
chemical potential. The standard BCS variational wave function is
given by
\begin{equation}
   | \Phi \rangle \ = \
   \mathcal{N} \
   \exp \left ( \sum_{\mathbf{k}} \ \frac{v_{\mathbf{k}}}{u_{\mathbf{k}}} \
   a^{\dag}_{\mathbf{k} \uparrow} a^{\dag}_{-\mathbf{k} \downarrow}
   \right ) \ | 0 \rangle
   \>,
\end{equation}
and it smoothly interpolates from the BCS to BEC limits, giving an
accurate description of the crossover. Here, the BCS parameters
$u_{\mathbf{k}}$, $v_{\mathbf{k}}$ only depend on $k\equiv
|\mathbf{k}|$ in the s-wave approximation, and $\mathcal{N} =
\prod_{\mathbf{k}} u_{\mathbf{k}}$ is the normalization constant,
such that $\langle \Phi | \Phi \rangle = 1$. With this ansatz, the
free energy becomes
\begin{align}
   \mathbf{F} \ = \ &
   2 \ \sum_{\mathbf{k}} \
   ( \epsilon_{\mathbf{k}} - \mu ) \ \rho_{\mathbf{k}}
   \ + \ \sum_{\mathbf{k,p}} \
   V_{\mathbf{p}-\mathbf{k}} \ \kappa^*_{\mathbf{k}} \kappa_{\mathbf{p}}
   \>,
\end{align}
where the \emph{normal} and \emph{anomalous} densities are
\begin{equation}
   \rho_{\mathbf{k}}
   \ = \
      \expect{ \Ad{\mathbf{k}\uparrow} \A{\mathbf{k}\uparrow} }
   \ = \ |v_{\mathbf{k}}|^2
   \>,
\end{equation}
and
\begin{equation}
   \kappa_{\mathbf{k}}
   \ = \
      \expect{ \A{-\mathbf{k}\downarrow} \A{\mathbf{k}\uparrow} }
   \ = \ v_{\mathbf{k}}^* u_{\mathbf{k}}
   \>,
\end{equation}
respectively, subject to the normalization condition,
\begin{equation}
   | u_{\mathbf{k}}|^2 \ + \ |v_{\mathbf{k}}|^2 \ = \ 1
   \>.
\end{equation}
Without loss of generality we can choose $u_{\mathbf{k}}$ and
$v_{\mathbf{k}}$ to be real.

After minimization, the resulting equations we have to solve are
\begin{align}
   \Delta_{\mathbf{k}}
   \ = \ &
   - \
   \sum_{\mathbf{p}} \
   V_{\mathbf{k}-\mathbf{p}} \
   \frac{\Delta_{\mathbf{p}}}{E_{\mathbf{p}}}
   \>,
\label{eq:gap}
   \\
   E_{\mathbf{k}}^2 \ = \ &
   (\epsilon_{\mathbf{k}} - \mu)^2 \ + \
   \Delta_{\mathbf{k}}^2
   \>,
   \\
   v_{\mathbf{k}}^2
   \ = \ &
   \frac{1}{2} \ - \
   \frac{\epsilon_{\mathbf{k}} - \mu}{2 E_{\mathbf{k}}}
   \>,
   \\
   u_{\mathbf{k}} v_{\mathbf{k}}
   \ = \ &
   \frac{\Delta_{\mathbf{k}}}{2 E_{\mathbf{k}}}
   \>,
\end{align}
with the added constraint that the total density $N$ of particles
is constant
\begin{equation}
   N \ = \ \langle \Phi | \hat{N} | \Phi \rangle
   \ = \ 2 \ \int \ \frac{\mathrm{d^3}k}{(2\pi)^3}  \ \rho_{\mathbf{k}}
   \>.
\label{eq:no}
\end{equation}

%
%

\section{Wave functions and excitation spectra}
\label{spectra}

\begin{figure}[h!]
   \includegraphics[width=0.48\textwidth]{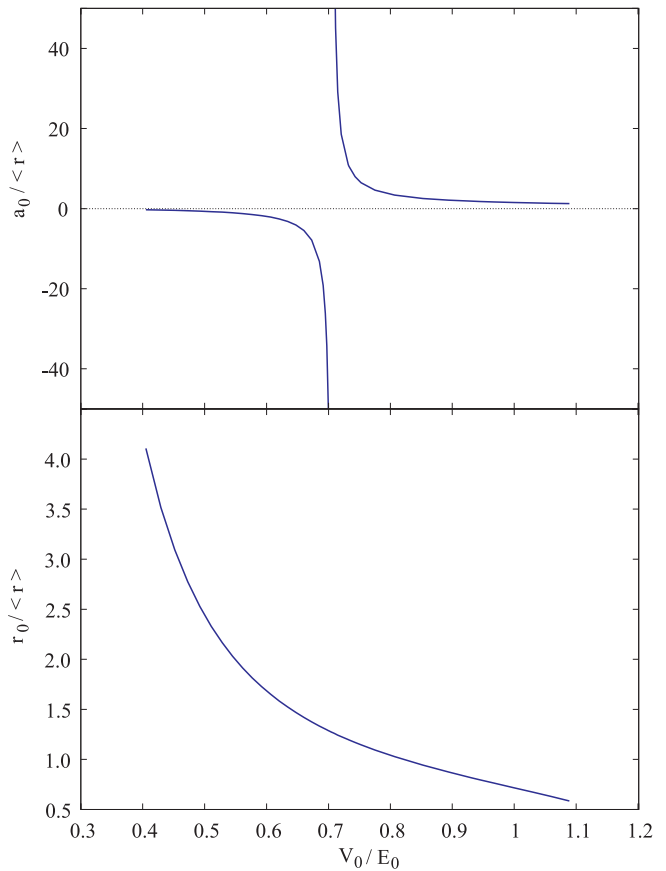}
   \caption{\label{fig1}
   (Color online)
   Scattering length~$a_0$  
   and effective range
   of the interaction $r_0$ 
   as a function of the
   potential depth $V_0/E_0$, where the energy scale
   $E_0=\hbar^2 / ( m \langle r \rangle^2)$.
   }
\end{figure}

\begin{figure}[h!]
   \includegraphics[width=0.48\textwidth]{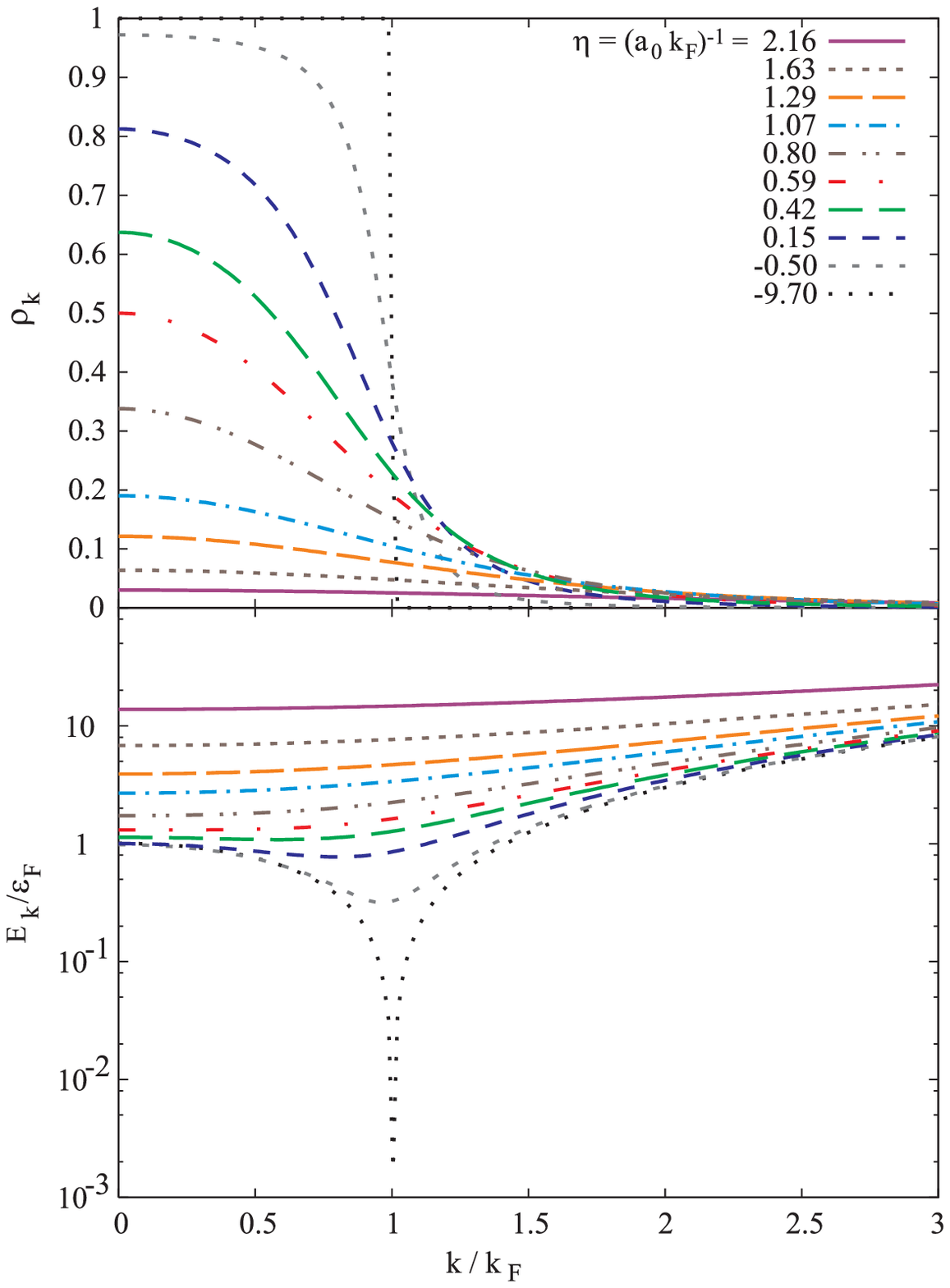}
   \caption{\label{fig2}
   (Color online)
   Momentum distributions $\rho_k$ (top) and the
   quasi-particle excitation spectra $E_k/\varepsilon_F$ (bottom)
   as a function of $k\equiv |\mathbf{k}|$,
   at fixed density ($k_F$) and various potential scattering lengths~$a_0$,
   where $\varepsilon_F$ is the Fermi energy and
   $k_F \langle r \rangle \approx~0.37$.}
\end{figure}

Numerical solutions are found for an attractive short-range
interaction, described by a Gaussian potential $V(r) = V_0 \exp(- b
r^2)$.
We explore the BCS-BEC crossover by fixing the width of the
potential $\langle r \rangle = 2/\sqrt{\pi b}$ and varying the
potential depth $V_0$, or implicitly the scattering length~$a_0$
of the interaction. Fig.~\ref{fig1} depicts all the essential
two-body physics of the problem. As the depth of the potential is
increased from zero, the scattering length
 grows, diverging to negative infinity at the first appearance of
a two-body bound state. On the bound-state side of the Feshbach
resonance the scattering length re-emerges from positive infinity.
Note that this procedure does not keep the \emph{effective} range
$r_0$ of the potential fixed. In the low-energy limit,
the effective range $r_0$, like the scattering length~$a_0$, can
be determined from the phase shift
\begin{equation}
   k \ \cot \delta_0 \ = \
   - \, \frac{1}{a_0} \ + \ \frac{1}{2} \, r_0 \ k^2
   \>,
\end{equation}
and is shown in Fig.~\ref{fig1}. The variation of the effective
range is small over the region of interest.

From now on, we shall focus on finite densities and the
crossover to BCS that can occur in this regime.
Like previous theoretical studies \cite{ref:sch_noz}, 
we find that the momentum distribution $\rho_k$ at constant density
smoothly evolves from the wave function of a molecule in momentum
space at large positive $(a_0 k_F)^{-1}$ to a Fermi-like
distribution at large negative $(a_0 k_F)^{-1}$, as depicted in
Fig.~\ref{fig2}~(top). During this transition, we see from
Fig.~\ref{fig2}~(bottom) that the quasi-particle excitation
spectrum $E_k$ develops a minimum at finite momentum,
signifying the BCS-BEC crossover point defined previously. Because the
density is non-zero, though small, the crossover point occurs for
positive $a_0 k_F$, where a two-body molecular bound state has
already formed.

\begin{figure}[h!]
   \includegraphics[width=0.48\textwidth]{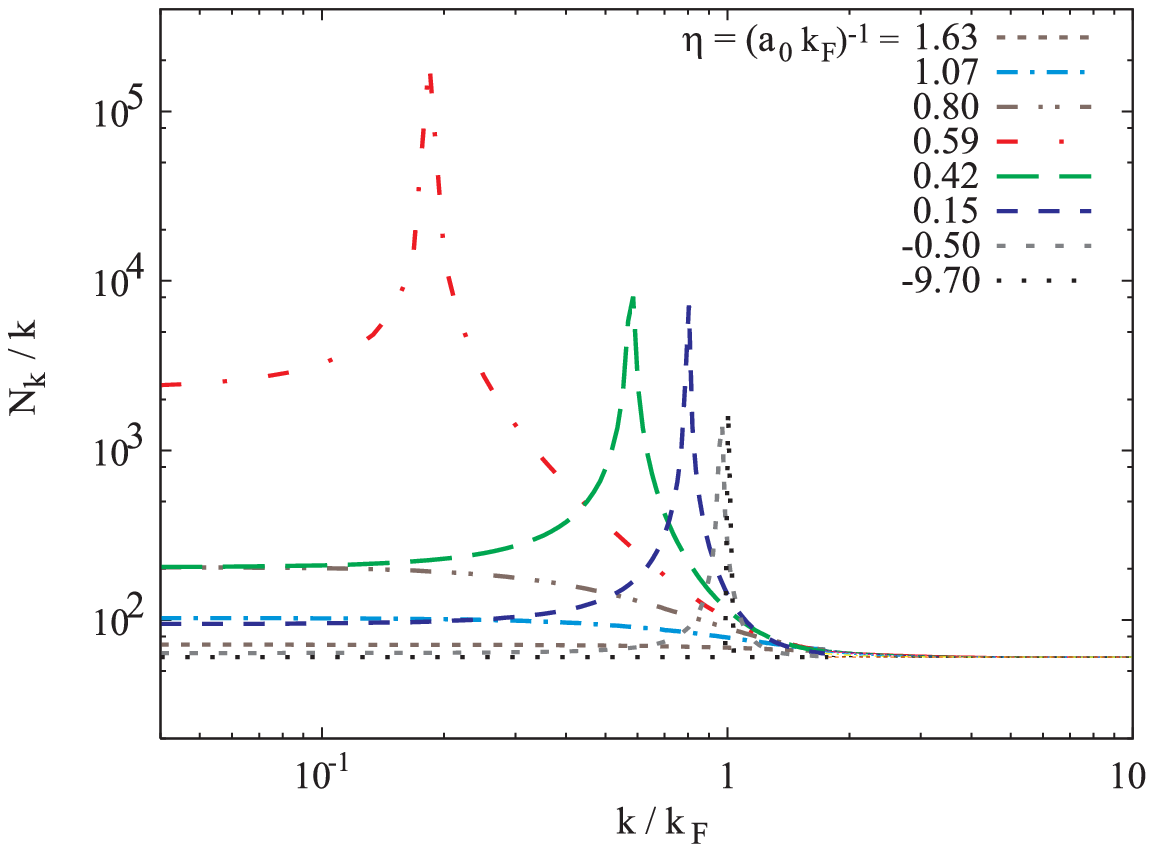}
   \caption{\label{fig3}
   (Color online)
   Density of states $N_k$ (as a function of momentum),
   at fixed $k_F$ and various potential scattering lengths~$a_0$,
   where $k_F \langle r \rangle \approx~0.37$.
   }
\end{figure}

The appearance of the minimum in the quasi-particle excitation
spectrum at non-zero momentum is clearly evident in the
quasi-particle density of states. As a function of momentum, the
density of states is defined as
\begin{equation}
   N_k = \frac{1}{\pi^2}
   \frac{k^2}{\left | \frac{\mathrm{d}E_k}{\mathrm{d}k} \right |}.
\end{equation}
Thus, $N_k$ is singular whenever there is a stationary point in
the excitation spectrum ($\mathrm{d}E_k/\mathrm{d}k = 0$), except
when $k=0$, where $N_k$ vanishes provided
$\mathrm{d}E_k/\mathrm{d}k \rightarrow 0$ no faster than $k^2$.
Fig.~\ref{fig3} shows how $N_k$ varies with scattering length at
constant particle density. The energy minimum for $k>0$ shows up
as a spike that migrates away from $k=0$ towards $k=k_F$ as $(a_0
k_F)^{-1}~\rightarrow~-\infty$, but the spike disappears when the
energy minimum occurs at $k=0$. Thus, the BCS-BEC crossover is the
point at which a spike first appears as we decrease $(a_0
k_F)^{-1}$.

\begin{figure}[t!]
   \includegraphics[width=0.48\textwidth]{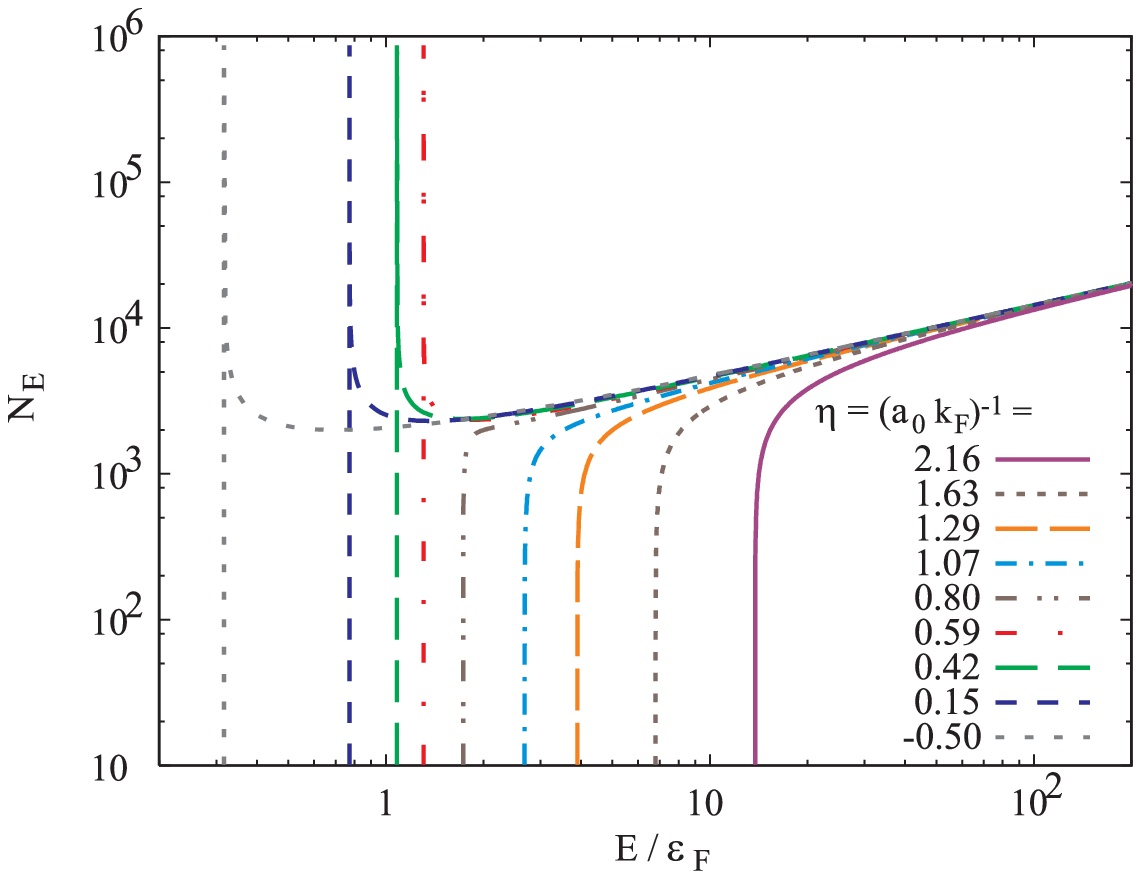}
   \caption{\label{fig4}
   (Color online)
   Density of states $N_E$ (as a function of energy),
   at fixed $k_F$, where $k_F \langle r \rangle \approx~0.37$,
   and various potential scattering lengths~$a_0$.
   }
\end{figure}

A more straightforwardly measurable quantity may be the density of
states as a function of energy
\begin{equation}
   N_E = \frac{1}{\pi^2}
   \frac{k_E^2}{\left | \frac{\mathrm{d}E_k}{\mathrm{d}k} \right |_E}
   \>,
\end{equation}
which is depicted in Fig.~\ref{fig4}.  Approximating the gap
parameter $\Delta_k$ as a constant, the BCS regime has a
singularity at finite momentum ($k = \sqrt{2m\mu}/\hbar$) of the
form
\begin{equation}
   N_E \simeq \frac{A}{\sqrt{E-\Delta}}
   \>,
\end{equation}
where the amplitude is
\begin{equation}
   A = \frac{1}{\pi^2 \hbar^3}\sqrt{\frac{m^3 \mu}{\Delta}}
   \>.
\end{equation}
We see that the size of the singularity will increase the further
we enter the BCS phase.

In the BEC limit, the density of states is zero at the energy
minimum occurring at $k=0$, and deep within the BEC phase it has
the following form close to the minimum
\begin{equation}
   N_E \simeq \frac{\sqrt{2m^3}}{\pi^2\hbar^3}
   \sqrt{E-|\mu|}
   \>.
\end{equation}

\begin{figure}[h!]
   \includegraphics[width=0.48\textwidth]{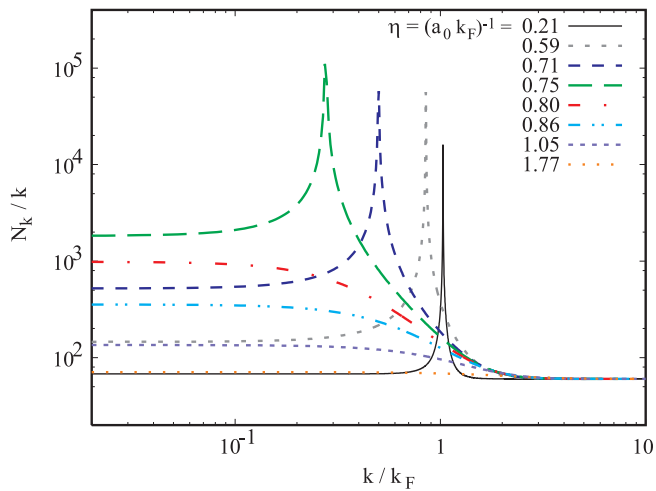}
   \caption{\label{fig5}
   (Color online)
   Density of states $N_k$ (as a function of momentum),
   at fixed potential scattering length
   ($a_0/\langle r \rangle~=~2.4$), and various values of $k_F$.
   }
\end{figure}

As well as considering the BCS-BEC crossover at constant density
arranged by tuning the interaction, we also examine the
\emph{density-driven} crossover at fixed interaction.
Fig.~\ref{fig5} illustrates the density of states as a function of
momentum for various particle densities, where we have chosen
$a_0$ to be positive so that a molecular bound state exists. We
observe the same disappearance of the spike in $N_k$ as $(a_0
k_F)^{-1} \rightarrow +\infty$ but, unlike Fig.~\ref{fig3}, the
extreme BCS limit
occurs when $(a_0 k_F)^{-1} \rightarrow 0$ since this corresponds
to infinite density.  Note also that both
Fig.~\ref{fig3}~and~\ref{fig5} show $N_k/k \rightarrow$ {const},
as $(a_0 k_F)^{-1} \rightarrow +\infty$, so that the pair-breaking
excitation spectrum has $(E_k -\delta E) \propto k^2$, where
$\delta E$ is the binding energy of the molecule. This is exactly
what we would expect for a non-interacting gas of Bose molecules.

\begin{figure}[t!]
   \includegraphics[width=0.48\textwidth]{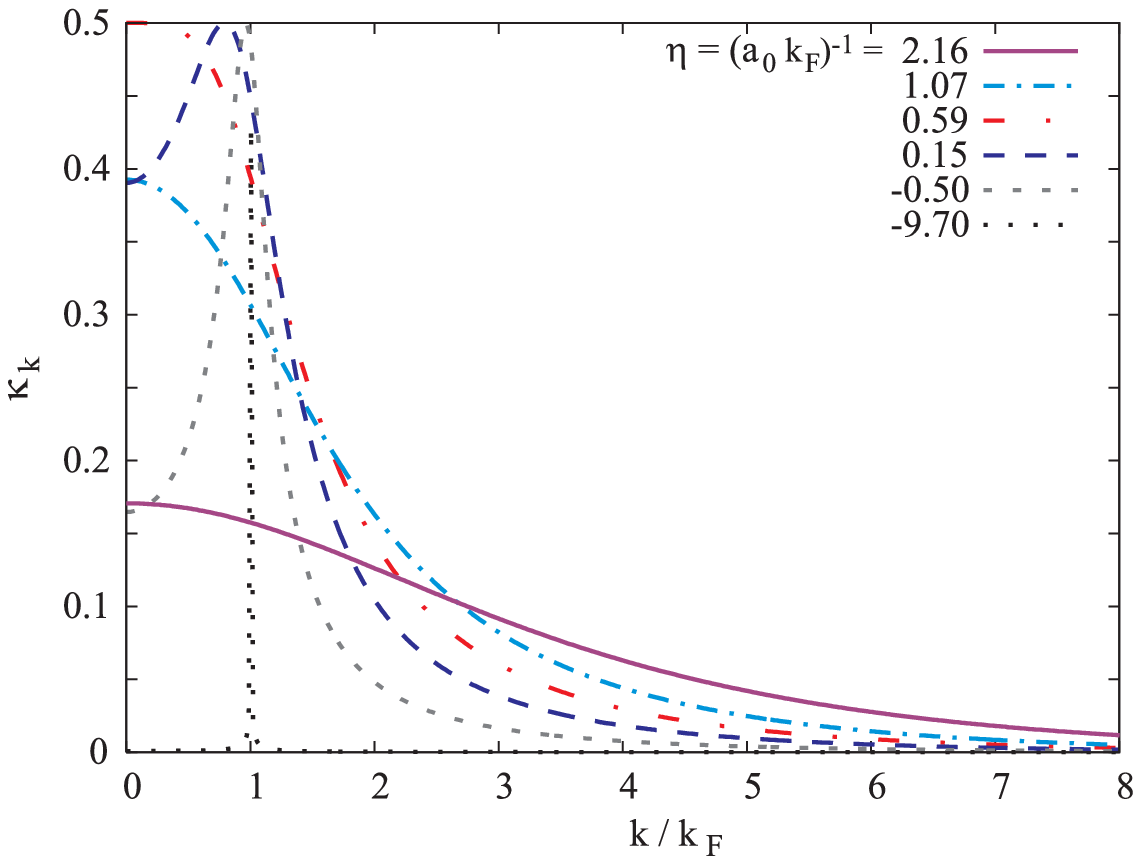}
   \caption{\label{fig6}
   (Color online)
   Momentum distribution of the condensate wave function,
   at fixed density ($k_F$) and various potential scattering lengths~$a_0$,
   where $k_F \langle r \rangle \approx~0.37$ .}
\end{figure}

Finally, Fig.~\ref{fig6} shows the behavior of the condensate wave
function throughout the crossover. 
Here we see the most pronounced features of the evolution from a
molecular state to the BCS limit where pairing exists only on the
Fermi surface.

%
%

\section{Phase diagram and comparison with zero-range potential}
\label{universal}

\begin{figure}[t!]
   \includegraphics[width=0.48\textwidth]{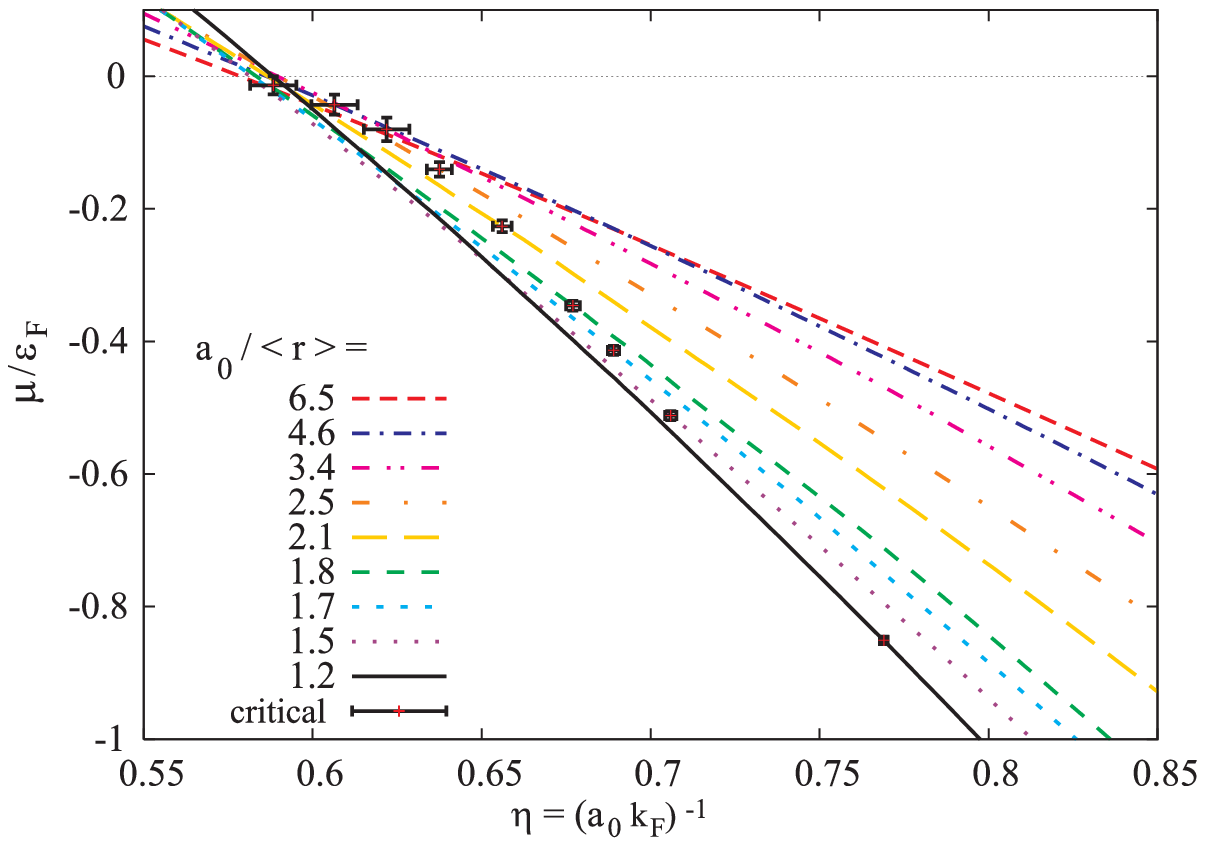}
   \caption{\label{fig:fig7}
   Profiles of the chemical potential $\mu/\varepsilon_F$ as a function of
   $\eta = (a_0 k_F)^{-1}$. The points mark the BEC-BCS
   crossover.
   }
\end{figure}

A more complete understanding of the BCS-BEC crossover can be
gained from considering the phase diagram for our finite-range
potential.
Figure~\ref{fig:fig7} shows the variation of chemical potential
$\mu/\varepsilon_F$ as a function of $(a_0 k_F)^{-1}$, for a set
of different scattering lengths~$a_0/\langle r \rangle$.
On each curve we mark the position of the BCS-BEC crossover point,
 which we refer to as the ``critical'' pair ($\mu_c$,
$k_{F\,c}$). The crossover point always occurs for $a_0>0$ and there is
no density-driven crossover when $a_0<0$. 
Since the size of $a_0$ dictates the size of the two-body 
molecular bound state while $a_0>0$, the crossover point in Fig.~\ref{fig:fig7}
is driven to lower densities when $a_0$ is increased.
As the zero-density limit ($k_{F\,c}$=0) is approached, the critical 
chemical potential $\mu_c/\varepsilon_F \rightarrow 0$, 
which is equivalent to the BCS-BEC crossover for a single-parameter potential 
initially discussed by Leggett.  However, as $k_{F\,c}$ increases, the
crossover point shifts to negative values of $\mu_c$ and 
larger $(a_0 k_F)^{-1}$.

It is useful to compare the results of our calculations with
previous studies of the BCS-BEC crossover that involve
single-parameter potentials. In the dilute or low-energy limit,
where $k_F \langle r \rangle, \langle r \rangle/a_0 \ll 1$, the
effective interaction $U$ can be written as
\begin{equation}
   U \ = \
   \frac{4\pi a_0}{m}
   \>,
\end{equation}
and it is independent of the details of the real interaction.
Substituting $U$ for the potential and eliminating high energies,
Eqs.~\eqref{eq:gap} and \eqref{eq:no} become, respectively
\begin{align}
   \int_0^\infty \ \mathrm{d} \varepsilon \
   \sqrt \varepsilon \
   \biggl [
      \frac{1}{\varepsilon} \ - \
      \frac{1}{\sqrt{ (\varepsilon - \tilde \mu)^2 + \tilde
      \Delta}}
   \biggr ]
   \ = \
   \pi \ \eta
   \>,
   \\
   \int_0^\infty \ \mathrm{d} \varepsilon \
   \sqrt \varepsilon \
   \biggl [
      1 \ - \
      \frac{\varepsilon - \tilde \mu}{\sqrt{ (\varepsilon -
      \tilde \mu)^2 + \tilde
      \Delta}}
   \biggr ]
   \ = \
   \frac{4}{3}
   \>,
\end{align}
where we have introduced the notations $\tilde \mu = \mu /
\varepsilon_F$, $\tilde \Delta = \Delta / \varepsilon_F$, and
$\eta = (a_0 k_F)^{-1}$. The above equations, first studied by
Leggett~\cite{ref:leg80}, are then solved for $\tilde \mu(\eta)$
and $\tilde \Delta(\eta)$. For obvious reasons, we will refer to
these functions as ``model independent'' or ``universal.''

\begin{figure}[t!]
   \includegraphics[width=0.48\textwidth]{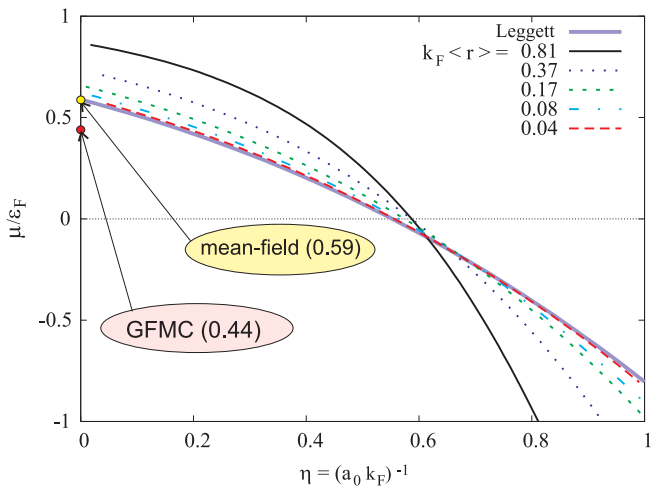}
   \caption{\label{fig8}
   (Color online)
   Comparison with Leggett's ``universal'' curve:
   We keep $k_F$ constant,
   and vary the scattering length~$a_0$.
   }
\end{figure}

We first compare our results for $\tilde \mu(\eta)$ against
Leggett's predictions for the case where the scattering
length~$a_0$ is varied and the density is held fixed, 
as shown in Fig~\ref{fig8}.  On the Leggett curve, the BCS-BEC crossover point 
($\mu=0$) occurs at $(a_0 k_F)^{-1} \approx 0.55$, and our numerical
$\tilde \mu(\eta)$ converges to this universal curve as $k_F
\langle r \rangle \rightarrow 0$.  Since typical experimental
parameters correspond to $k_F \langle r \rangle \simeq 0.04$, 
the ultracold atomic gases clearly lie on the Leggett curve and are
well described by a single-parameter potential.

The value of $\mu/\varepsilon_F$ for $\eta=0$ is usually referred
to as the ``unitarity'' limit~\cite{unitarity,bertsch,carlson}. In
this limit one expects that all sensitivity to the detailed
features of the interaction is lost~\cite{schiff}, and the system
energy is determined entirely by the density. As a result, this
limit is particularly sensitive to many-body correlation effects,
and the Leggett curve predicts this value as being 0.59~(see
also~\cite{ref:ERM97}). Recently, accurate calculations based on
the Green's Function Monte Carlo (GFMC) method~\cite{carlson},
have lowered the upper-bound on this result to ($0.44 \pm 0.01$),
which shows that beyond mean-field effects account for at least a
25\% improvement in the binding energy over the mean-field result.

\begin{figure}[t!]
   \includegraphics[width=0.48\textwidth]{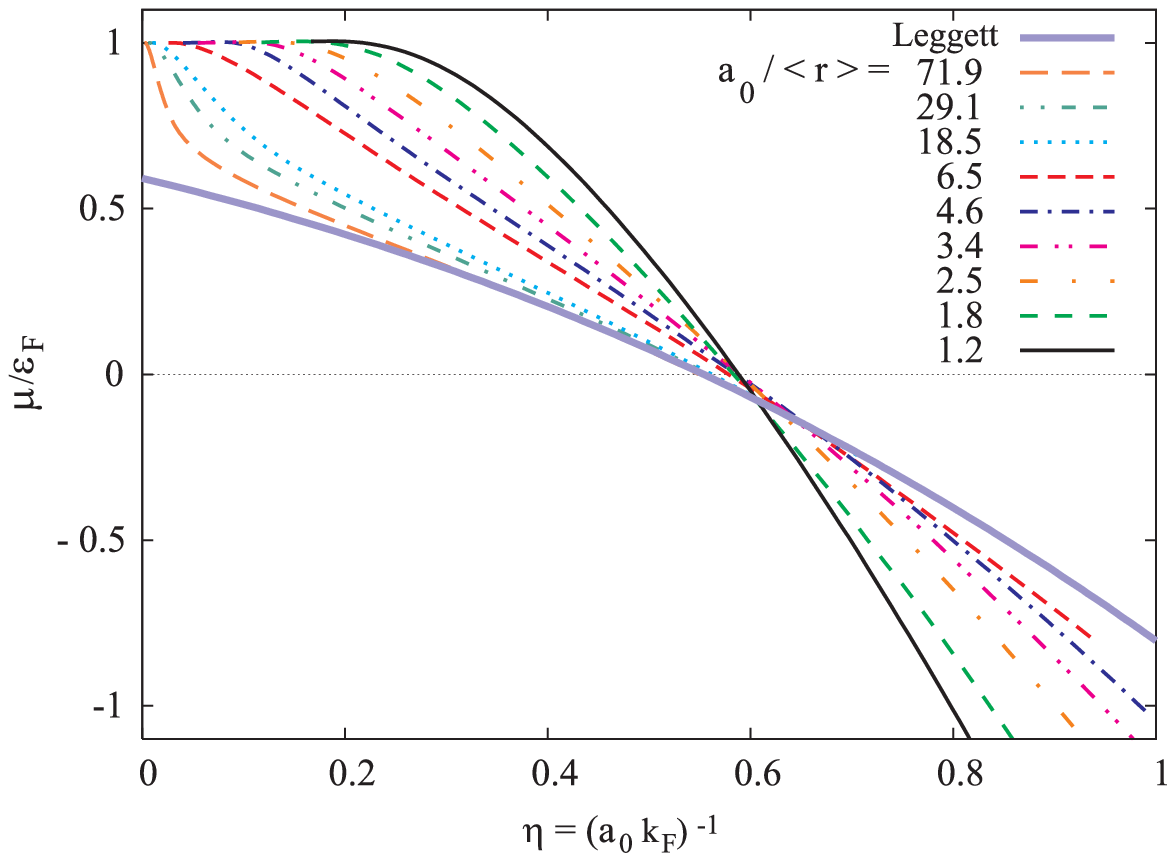}
   \caption{\label{fig9}
   (Color online)
   Comparison with Leggett's ``universal'' curve:
   The scattering length~$a_0$ is held constant,
   the density  ($k_F$) is varied.
   }
\end{figure}

Fig.~\ref{fig9} compares the numerical $\tilde \mu(\eta)$ with the
universal curve for the case where the density is varied and
$a_0/\langle r \rangle$ is fixed. Not surprisingly, the numerical
$\tilde \mu(\eta)$ approaches the universal curve as scattering
length increases, and we require $a_0/\langle r \rangle > 18$
before $\tilde \mu(\eta)$ touches the universal curve.
Furthermore, we see that no matter how large we make $a_0$, the
universal curve breaks down in the region sufficiently close to
$(a_0 k_F)^{-1}=0$.  This point corresponds to infinite density,
the BCS limit in the density-driven crossover, so we must always
have $\tilde \mu(0) = 1$, which is consistent with our numerical
results.  If we consider $\tilde \mu(\eta)$ for which $a_0/\langle
r \rangle \approx 72$ we observe that it collapses onto the
universal curve when $(a_0 k_F)^{-1}>0.3$.  This translates into
the condition $k_F \langle r \rangle < 0.046$ for the universal
equations to be valid, which is consistent with Fig.~\ref{fig8}.

%
%

\section{Conclusion}
\label{conclusions}

In this paper we have performed a study of the BCS-BEC crossover,
at zero temperature, in the single-channel model, using a
realistic effective interaction. In the mean-field approximation
the crossover is described using a variational wave function,
which interpolates smoothly between the BEC and BCS limits.  
We find that the crossover point always lies on the
side of the Feshbach resonance where a molecular bound state
exists at zero density, and that it converges to that of a
single-parameter potential in the low density limit.
Moreover, our results reproduce the usual Leggett picture in the 
limit of a dilute Fermi gas, as expected. 
By using a potential of finite range, we have shown that 
the crossover for typical experimental parameters in ultracold atomic 
Fermi gases can be described by a zero-range, single-parameter potential.
We emphasize that a clear signature of the crossover point
appears in the density of states, which could be measured
spectroscopically~\cite{ref:TZ00}.

The mean-field description of the single-channel model presented
here is not entirely satisfactory. As described above, recent GFMC
results provide an upper value for the energy density in the
dilute limit ($\eta=0$), which represents a 25\% improvement over
the Leggett limit. Therefore, additional work is required in order
to include next-to-leading order effects in a quantitative way,
and study the changes (or lack thereof) in the features of the
BEC-BCS crossover. Approximations schemes based on the
two-particle irreducible effective action~\cite{2PI} are readily
available in quantum field theory where they have been recently
employed to describe the dynamics of a system out of equilibrium,
and the dynamics of phase transitions~\cite{BVA}. Work is
currently under way in order to apply such methods to the study of
the BEC-BCS crossover based on the Hamiltonian~\eqref{eq:ham_0}.

The single-channel model is of course a simplification of detailed
models for the Feshbach resonance which include the resonance
explicitly, either as a boson (the Fermi-Bose
model~\cite{fermibose}) or more generally as three-level~
\cite{3level} or four-level fermi systems. The Fermi-Bose model,
as well as models with even numbers of fermions, all have large
regimes of parameter space where the resonance is sufficiently
detuned from the open channel that high energy degrees of freedom
can be integrated out in favor of an effective interaction between
two species, recovering the model discussed in this paper. Close
enough to resonance the detailed fermionic structure of the
resonance may become important. Models where the Feshbach bound
state shares a state with the open channel (relevant to $^{40}K$)
are different, and do not reduce to an effective single channel
model in a simple way\cite{3level}. In future work, we will
compare and contrast these different models~\cite{manymodel} 
using a generalization of the variational scheme given in this paper.

%
%
\begin{acknowledgments}
M.M.P. acknowledges support from the Association of Commonwealth
Universities and the Cambridge Commonwealth Trust.
\end{acknowledgments}

\vfill

%
%
%
%
%

%
%
\end{document}